\documentclass[aps,prb,twocolumn,superscriptaddress]{revtex4-2}

\usepackage{amsmath}
\usepackage{mathtools} 
\usepackage{amssymb}
\usepackage{amsthm}
\usepackage{bm} 
\usepackage{upgreek} 
\usepackage{xcolor}
\usepackage{graphicx}
\usepackage{epstopdf}
\usepackage[colorlinks,linkcolor=blue,anchorcolor=blue,citecolor=blue,urlcolor=blue]{hyperref}

\newcommand{\nn}{{\nonumber}}
\newcommand{\bea}{\begin{eqnarray}}
	\newcommand{\eea}{\end{eqnarray}}
\newcommand{\ie}{\textit{i.e.{ }}}

\newcommand{\md}{\mathrm{d}}

\begin{document}
\title{Magnetotransport in overdoped La$_{2-x}$Sr$_x$CuO$_4$: Effect of anisotropic scattering}

\author{Rui-Ying Mao}
\affiliation{National Laboratory of Solid State Microstructures $\&$ School of Physics, Nanjing University, Nanjing 210093, China}

\author{Da Wang} \email{dawang@nju.edu.cn}
\affiliation{National Laboratory of Solid State Microstructures $\&$ School of Physics, Nanjing University, Nanjing 210093, China}
\affiliation{Collaborative Innovation Center of Advanced Microstructures, Nanjing University, Nanjing 210093, China}

\author{Qiang-Hua Wang} \email{qhwang@nju.edu.cn}
\affiliation{National Laboratory of Solid State Microstructures $\&$ School of Physics, Nanjing University, Nanjing 210093, China}
\affiliation{Collaborative Innovation Center of Advanced Microstructures, Nanjing University, Nanjing 210093, China}

\begin{abstract}
We revisit the Hall effect and magnetoresistivity by incorporating the anisotropic scattering caused by apical oxygen vacancies in overdoped La-based cuprates. The theoretical calculations within the Fermi liquid picture agree well with a handful of anomalous magneto-transport data, better than the results using an isotropic scattering rate alone. In particular, we obtain the upturn of Hall coefficient $R_H$ with decreasing temperature $T$, the initial drop of $R_H$ in magnetic field $B$ in all overdoped regimes, the linear resistivity $\rho$ versus $B$ near the van Hove doping level, the temperature dependence of the magnetoresistivity ratio, and the violation of Kohler's law. These results suggest that many of the anomalous transport behaviors in overdoped La$_{2-x}$Sr$_x$CuO$_4$ could actually be understood within the Fermi liquid picture.
\end{abstract}
\maketitle

\section{INTRODUCTION}

As the first family of high temperature superconductors discovered so far, cuprates continue to challenge our understanding of many body physics, in particular the properties of a doped Mott insulator \cite{Lee2006}. However, it is hoped that at least in the overdoped regime, the correlation effect might be weakened relatively, such that a Fermi liquid picture could be applied. On this basis, the superconducting state could also be described properly by the Bardeen-Cooper-Schrieffer (BCS) theory. If this were the case, we would have a good starting point to descriminate what falls within/beyond the Fermi liquid theory at lower doping levels \cite{keimer2015quantum}. The crossover doping level above which the Fermi liquid picture applies is not yet clear, and in fact, many anomalous phenomena reported in recent expeiments seem to push the ``boundary'' of the Fermi liquid phase to very high doping levels, even beyond the superconducting domes \cite{Bozovic2019}. To gain further insights, it is important to investigate the anomalies more closely to see whether and how they could actually be described within the Fermi liquid picture.

The upturn of the Hall coefficient $R_H$ with decreasing temperature $T$ in overdoped La$_{2-x}$Sr$_x$CuO$_4$ (LSCO) \cite{Hwang1994} is an early signature of the breakdown of the Fermi liquid theory, since within the latter picture $R_H$ should be $T$-independent unless the system has both electron-like and hole-like Fermi pockets. The upturn of $R_H$ was later confirmed to persist for all doping levels up to $x=0.36$ \cite{Tsukada2006}. In particular, at $x=0.32$, $R_H$ grows continuously from negative to positive values with decreasing temperature.
Another signature of the Fermi liquid breakdown comes from the linear resistivity under strong magnetic fields that recovers the normal states from the superconducting state. The $T$-linear resistivity at low temperatures persists up to $x\approx 0.3$ in LSCO \cite{Cooper2009}, which is surprising since the doping level is readily far from the hypothetic quantum critical point at $x_c\approx 0.2$. Subsequent careful analysis of the magnetoresistivity data at $x=0.23$ reveals a special quadratic scaling \cite{Berben2022} consistent with the phenomenological ``quadrature Planckian'' dissipation $\tau^{-1} = \sqrt{T^2+\gamma^2B^2}$ \cite{Ayres2021}. This includes an unusual resistivity linear in $B$ at low temperatures.\cite{Giraldo-Gallo2018} In LSCO, the Fermi surface changes from being hole-like to electron-like as the doping level increases through the van Hove singularity (VHS) at $x_{\rm vhs}=0.2$. One expects a corresponding change in the Hall coefficient. Indeed, in the strong field limit, $R_H=1/(1+x)e$ for $x<x_{\rm vhs}$, and $-1/(1-x)e$ for $x>x_{\rm vhs}$, respectively\cite{Lifshitz1957}. However, in the weak field limit, $R_H$ is found to smoothly decrease with doping, from positive to negative values, and changes sign at $x_0\approx0.3$\cite{Tsukada2006}. Recently, these behaviors are qualitatively reproduced within the Fermi liquid picture under an isotropic scattering rate \cite{Mao2021}.

On the other hand, in the superconducting state of overdoped LSCO films, the low temperature superfluid density $\rho_s$ is found to decrease linearly with $T$ \cite{Bozovic2016}, which is the expected behavior of a clean d-wave superconductor \cite{Hirschfeld1993}. However, the zero temperature superfluid density $\rho_s(0)$ is found to scale linearly with the transition temperature $T_c$, which is instead a typical behavior of a dirty BCS superconductor \cite{AG1961}. This dilemma is reconciled only recently by the recognition of the unique property of apical oxygen vacancies \cite{Wang2022}. {Such impurities are known to be more populated with over-doping \cite{Kim2017}}. In the Born limit, they cause an anisotropic scattering rate $\Gamma_d\cos^2(2\theta)$ (with $\theta$ the azimuthal angle relative to the anti-nodal direction) {through second order hopping processes connecting the nearest neighboring sites} \cite{Wang2022}. The global scale of the scattering rate affects the transition temperature, hence can produce the behavior $\rho_s(0)\propto T_c$ in the dirty limit. The depletion of the superfluid density at low temperatures are contributed by nodal quasiparticle excitations, but since the above form of scattering rate vanishes in the nodal direction, the behavior is similar to what would be found in the clean limit, even if the global scale sets the system in the dirty limit. This solves the puzzle regarding the two linear scaling behaviors of the superfluid density\cite{Wang2022}. Furthermore, such an anisotropic scattering suggests that the optical conductivity (versus the frequency) is an integration of Lorentzians over a distribution of scattering rates, hence becomes increasingly sharp as the frequency approaches zero. Therefore a single-mode Lorentzian fit of the optical conductivity data at finite frequency would under-estimate the Drude weight.\cite{Wang2022} This resolves the so-called ``missing'' of Drude weight upon overdoping \cite{Mahmood2019}.

In view of the important effect of the anisotropic scattering rate, we are motivated to re-examine the magnetotransport in overdoped LSCO, so far only considered theoretically in the isotropic scattering limit \cite{Mao2021}. We find that incorporating the anisotropic scattering rate further provides better agreement with a handful of experiments, such as the the upturn of $R_H$ with decreasing $T$ \cite{Hwang1994, Tsukada2006}, the initial drop of $R_H$ in $B$ for all overdopings \cite{Balakirev2009, Collignon2017}, the linear resistivity $\rho$ versus $B$\cite{Cooper2009, Giraldo-Gallo2018}, the temperature dependence of the magnetoresistivity ratio $[\rho(B)-\rho(0)]/\rho(0)$ \cite{Kimura1996,Vanacken2005,Cooper2009,Giraldo-Gallo2018}, and the violation of Kohler's law \cite{Kimura1996,Vanacken2005}. These results suggest that many of the anomalous transport behaviors in overdoped LSCO could actually be understood within the Fermi liquid picture.

\section{MODEL AND METHODS}

\begin{figure}
\includegraphics[width=\linewidth]{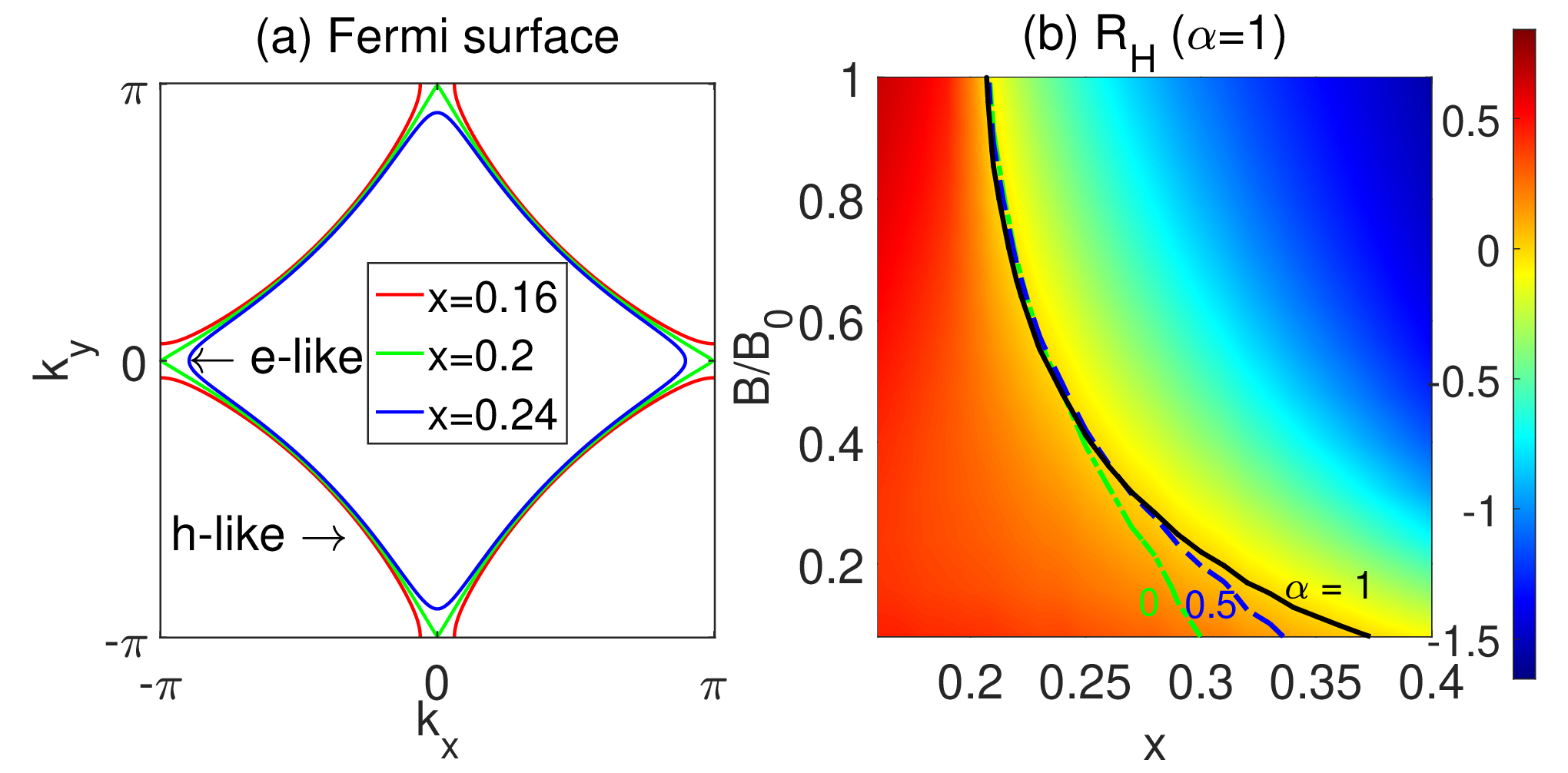}
\caption{ (a) Fermi surface for three doping levels near the VHS doping $x_{\rm vhs}$. (b) $R_H$ as a function of $x$ and $B$. The color encodes the value of $R_H$ for $\alpha=1$, and the black curve is the contour across which $R_H$ changes sign. The same contour for $\alpha=0.5$ (dashed blue line) and $0$ (dashed-dotted green line) are also shown for comparison.}
\label{fig:fs-rh}
\end{figure}

The band structure of overdoped LSCO as obtained from  angle-resolved photoemission spectra (ARPES) can be described quite well by a one band tight binding model defined on the square lattice with hoppings $t$, $t'=-0.12t$ and $t''=0.06t$ between the first, second and third nearest neighboring sites \cite{Yoshida2006,Horio2018}. For simplicity, we assume a purely two-dimensional rigid band structure, \ie only the chemical potential $\mu$ is to be tuned to match the doping level. In this model, there is a VHS at $x_{\rm vhs}\approx0.197$, across which the Fermi surface (FS) topology changes from hole- to electron-like, as shown in Fig.~\ref{fig:fs-rh}(a).

Throughout this work, the magnetic field $\0B$ is assumed to be perpendicular to the basal plane, which is the same as in most experiments. According to the Luttinger theorem \cite{Luttinger1960}, the enclosed area of the FS is $(1+x)S_{\rm BZ}/2$ ($\+S_{\rm BZ}$ the Brillouin zone area) for $x<x_{\rm vhs}$ and $(1-x)S_{\rm BZ}/2$ for $x>x_{\rm vhs}$, which gives rise to the Hall coefficient $R_H=1/(1+x)e$ and $-1/(1-x)e$, respectively, in the strong field limit $B\to\infty$ \cite{Lifshitz1957}. However, in the weak field limit, the Hall coefficient loses the above topological meaning (unless in the continuum limit or a band with quadratic dispersion) and instead should be determined by the Fermi surface curvature as clarified geometrically by Ong \cite{Ong1991}, or equivalently captured by the Kubo formula for the longitudinal conductivity
\begin{align}\label{eq:kubo1}
	\sigma_{xx}=\frac{2 e^2}{N_k} \sum_{\0k}\tau(\0k)v_{x}^2(\0k)\left(-\frac{\partial f}{\partial \varepsilon_\0k}\right) ,
\end{align}
where $2$ comes from spin, and the Hall conductivity
\begin{align}\label{eq:kubo2}
	\sigma_{xy}=-\frac{2 e^3B}{N_k} \sum_{\0k } \tau^2(\0k) v_x \left(v_x \frac{\partial v_y}{\partial k_y} - v_y \frac{\partial v_y}{\partial k_x}\right) \left(-\frac{\partial f}{\partial \varepsilon_\0k}\right) ,
\end{align}
where $v_{x,y}=\partial \varepsilon_\0k/\partial k_{x,y}$ and $f$ is the Fermi distribution function. The collision time $\tau(\0k)$ is the inverse of the total scattering rate given by
\begin{align}\label{eq:gamma(k)}
	\tau^{-1}(\0k)&=\Gamma_s+\frac{1}{4}\Gamma_d (\cos k_x-\cos k_y)^2,
\end{align}
where $\Gamma_s$ and $\Gamma_d$ are isotropic and anisotropic scattering amplitudes. We assume the $\Gamma_d$-term arises from the apical oxygen vacancies in overdoped LSCO \cite{Wang2022}. (Below the optimal doping level, the anisotropic scattering is also proposed in the phenomenological ``cold spot'' model \cite{Ioffe1998}. But here we will limit ourselves to over doping only, as it might be hopeless to apply the Fermi liquid theory below optimal doping, where the strong correlation effects beyond the Fermi liquid picture are known to be essential.) In the following, we introduce a dimensionless parameter $\alpha$ to quantify the anisotropy fraction such that $\Gamma_d=\alpha\Gamma_0$ and $\Gamma_s=(1-\alpha)\Gamma_0$, with $\Gamma_0=\Gamma_s+\Gamma_d$.
We note that in Eq.~\ref{eq:kubo2} the Hall conductivity $\sigma_{xy}$ is proportional to $e^3B$. This should be compared to the quantized Hall conductivity in unit of $e^2/h$ in two-dimensional quantum Hall systems or Chen insulators.

The Kubo formula is applicable in the limit of $B\rightarrow 0$. In order to go beyond the $B$-linear Hall conductivity, we adopt the Chambers' formula \cite{Chambers1952}
\begin{align}\label{eq:chambers}
	\sigma_{\alpha\beta}=&\frac{2e^3B}{(2\pi)^2}\int \md\varepsilon \left(-\frac{\partial f}{\partial \varepsilon}\right) \nn\\
	&\int_0^\+T\md t v_\alpha[\0k(t)] \int_{-\infty}^t\md t' v_\beta[\0k(t')] \exp\left\{-\int_{t'}^{t} \frac{\md s}{\tau[\0k(s)]}\right\},
\end{align}
which can be derived from the Boltzmann transport equation \cite{Budd1962, Abrikosov1988}. Within the semiclassical picture, the electrons are driven by the Lorentz force $\dot{\0k}=-e\0v_\0k\times\0B$ and move on cyclotron orbits in the momentum space. This semiclassical picture can be justified by the observation of cyclotron resonance in LSCO \cite{Post2021}. At low temperatures, the cyclotron orbit is limited on the Fermi surface (because of the derivative of the Fermi function in the above formula). We obtain the time-dependent momentum $\0k(t)$, velocity $\0v(t)$ and lifetime $\tau(t)$. Since both $\0v(t)$ and $\tau(t)$ are periodic functions with the cyclotron period $\+T$, the integral over $t'$ can be performed within each period, yielding
\begin{align} \label{eq:chambers2}	\sigma_{\alpha\beta}=&\frac{2e^3B}{(2\pi)^2} \left[1-\exp\left(-\int_0^{\+T}\frac{\md s}{\tau(s)}\right)\right]^{-1} \nn\\&\int_0^{\+T}\md t v_\alpha(t) \int_{t-\+T}^t\md t' v_\beta(t')
\exp\left[-\int_{t'}^{t}\frac{\md s}{\tau(s)}\right] ,
\end{align}
which is convenient for numerical calculations \cite{Maharaj2017}. Using Eq.~\ref{eq:chambers2} we calculate both $\sigma_{xx}$ and $\sigma_{xy}$. The conductivity tensor is then reversed to obtain the longitudinal resistivity $\rho_{xx}$ (denoted by $\rho$ for simplicity) and Hall coefficient $R_H$,
\begin{align}
\begin{bmatrix}
\rho & -R_HB \\ R_HB & \rho
\end{bmatrix} = \begin{bmatrix}
\sigma_{xx} & \sigma_{xy} \\ -\sigma_{xy} & \sigma_{xx}
\end{bmatrix}^{-1}.
\end{align}

To quantify the magnetic field, we define a characteristic magnetic field $B_0=m_c^*/e\tau$ with $\tau=1/\Gamma_0$ a time scale and $m_c^*=1/ta^2$ a mass scale ($a$ is the lattice constant). As a rough estimation, we have $B_0 \sim 100$T for LSCO\cite{Mao2021}. In the following, we use the Kubo formula for $B/B_0\to0$, while the results for finite $B/B_0 > 0.002$ are obtained by the Chambers formula. The latter can be applied for any nonzero $B$, but becomes numerically inefficient in the limit of $B/B_0\to 0$.

\section{RESULTS}

\subsection{Hall coefficient}
We first consider the doping and magnetic field dependence in $R_H$ (at low temperatures). The numerical result is shown in Fig.~\ref{fig:fs-rh}(b). The color scale shows the value of $R_H$ for a purely anisotropic scattering, $\alpha=1$. The solid black line highlights where $R_H$ changes sign. In the strong field limit, where $\+T\to 0$, $R_H$ changes abruptly from $1/(1+x)$ at $x<x_{\rm vhs}$ to $-1/(1-x)$ at $x>x_{\rm vhs}$. This behavior is independent of the scattering anisotropy, as the system is effectively in the ballistic regime, such that the Hall conductivity is purely determined by the Luttinger volume enclosed by the Fermi surface \cite{Lifshitz1957}.
As the field decreases, the cyclotron motion becomes slow and sensitive to the local scattering rate. As a result, the conductivity tensor becomes sensitive to the local curvature of the Fermi surface. The full cyclotron motion can be decomposed into segments.
Near the nodal point, the quasiparticle moves on a hole-like segment in our doping range, while near the antinodal point, the quasiparticle moves on electron-like segment for $x>x_{\rm vhs}$, see Fig.~\ref{fig:fs-rh}(a). The total conductivity tensor is given by a weighted sum of the segments. For the anisotropy fraction $\alpha=1$, the nodal quasiparticles experience vanishing scattering, hence make leading contribution to the conductivity. This explains why $R_H>0$ for all $x$ near and above $x_{\rm vhs}$. For even higher doping levels, the electron-like segment increases and eventually $R_H$ becomes negative. This explains the sign change of $R_H$ with increasing doping at low fields. For comparison, we also present the transition lines (where $R_H$ changes sign) for scattering anisotropy $\alpha =0.5$ (dashed blue line) and $0$ (dotted-dashed green line) in Fig.~\ref{fig:fs-rh}(b). We see the transition doping level is higher for larger scattering anisotropy. This is because a larger anisotropy enhances more strongly the relative contribution of the hole-like nodal quasiparticles to the conductivity tensor.

\begin{figure}
\includegraphics[width=\linewidth]{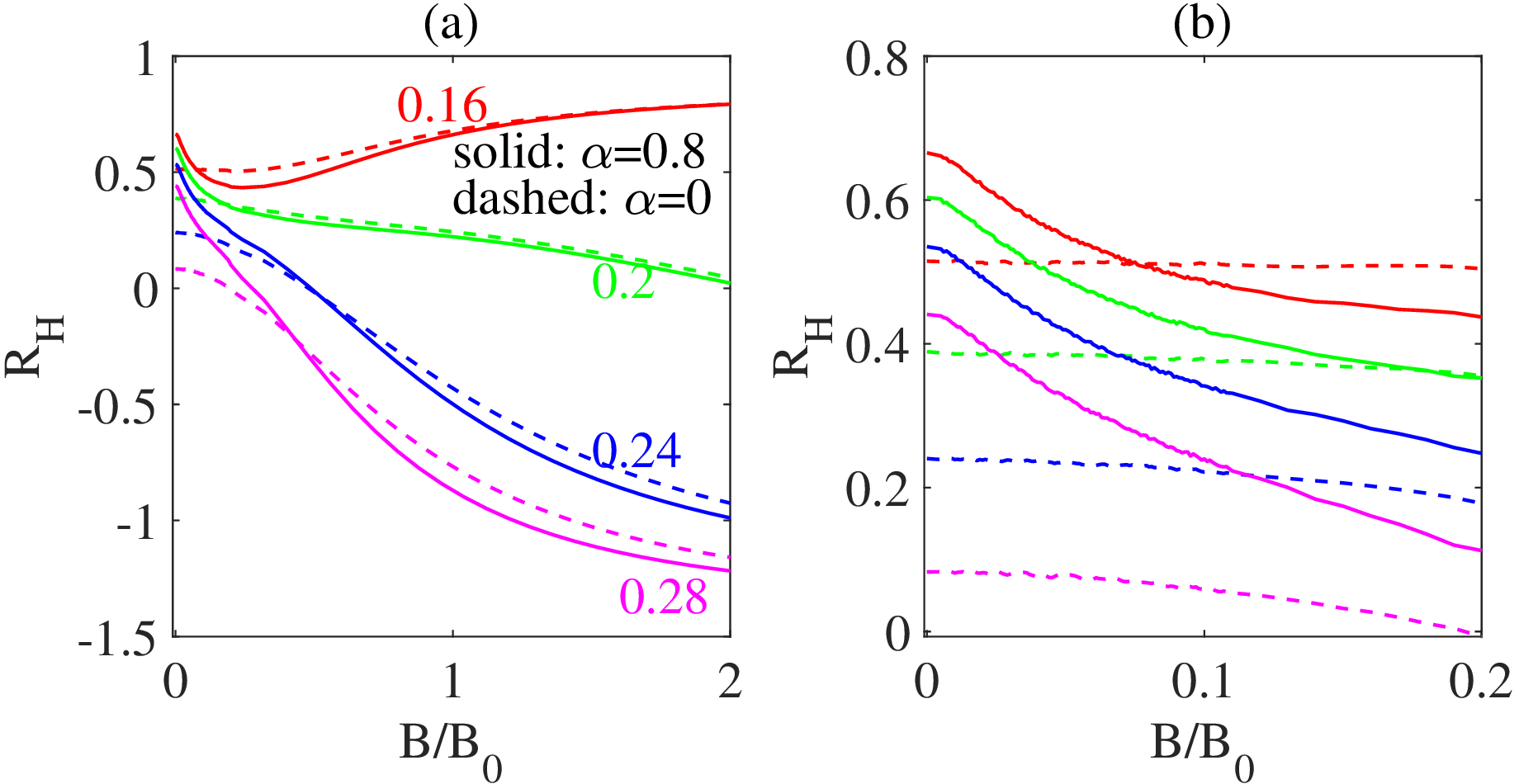}
\caption{Magnetic field dependence of $R_H$ in the case of $\alpha=0.8$ (solid lines) and $\alpha=0$ (dashed lines) at four different doping levels $x=0.16$, $0.2$, $0.24$, $0.28$. (b) is a zooming of (a) at low fields.}
\label{fig:rh-lambda}
\end{figure}

For moderate fields, $R_H$ should interpolate between the high and weak field limits. This is roughly the case in Fig.~\ref{fig:fs-rh}(b). Interestingly, closer examination of our results reveal an interesting structure in the evolution of $R_H$: it always drops with $B$ at small fields, and drops more quickly as the scattering anisotropy $\alpha$ increases, see Figs.~\ref{fig:rh-lambda}(a) and \ref{fig:rh-lambda}(b). This initial drop of $R_H$ in $B$ happens for all doping levels on the two sides of the VHS, and is consistent with the experiments \cite{Balakirev2009, Collignon2017} that appeared puzzling beforehand. To gain qualitative insights, we consider an effective ``two band'' model \cite{AshcroftBook}, in which the two bands should be understood as two types of segments near the nodal and antinodal points. In this simplified model, the total Hall coefficient can be calculated up to the $B^2$ term,
\begin{align}\label{eq:twobandmodle}
R_H&=\frac{\rho_{1}^2 R_2 + \rho_{2}^2 R_1+R_1 R_2(R_1+R_2)B^2}{(\rho_{1}+\rho_{2})^2+(R_1+R_2)^2 B^2}\nn\\ &\approx \frac{\rho_1^2R_2+\rho_2R_1^2}{(\rho_1+\rho_2)^2}- \frac{(R_1+R_2)(\rho_1R_2+\rho_2R_1)^2}{(\rho_1+\rho_2)^4}B^2,
\end{align}
where $R_{1,2}$ and $\rho_{1,2}$ are Hall coefficient and resistivity for the two bands, respectively. As long as $R_1+R_2>0$, which is clearly true for $x<x_{\rm vhs}$ since both $R_{1}$ and $R_{2}$ are positive, $R_H$ drops with $B$. While for $x>x_{\rm vhs}$, the hole-like segment is much larger than electron-like one in our range $x<0.4$, so we still have $R_1+R_2>0$. This qualitative picture of the initial drop of $R_H$ in $B$ is consistent with our numerical results shown in Fig.~\ref{fig:rh-lambda}(a). The quadratic dependence at small $B$ is dictated by time-reversal symmetry, and is also correctly reproduced in Fig.~\ref{fig:rh-lambda}(b).
Therefore, the initial drop of $R_H$ in $B$ is a universal feature in our doping range ($0.16<x<0.4$) on both sides of the VHS. Furthermore, by comparing the results of $\alpha=0$ and $\alpha=0.8$, see Fig.~\ref{fig:rh-lambda}, the anisotropic scattering is found to make a steeper initial drop (at finite but small $B$), in better agreement with the experiments \cite{Balakirev2009,Collignon2017}.

\begin{figure}
\includegraphics[width=\linewidth]{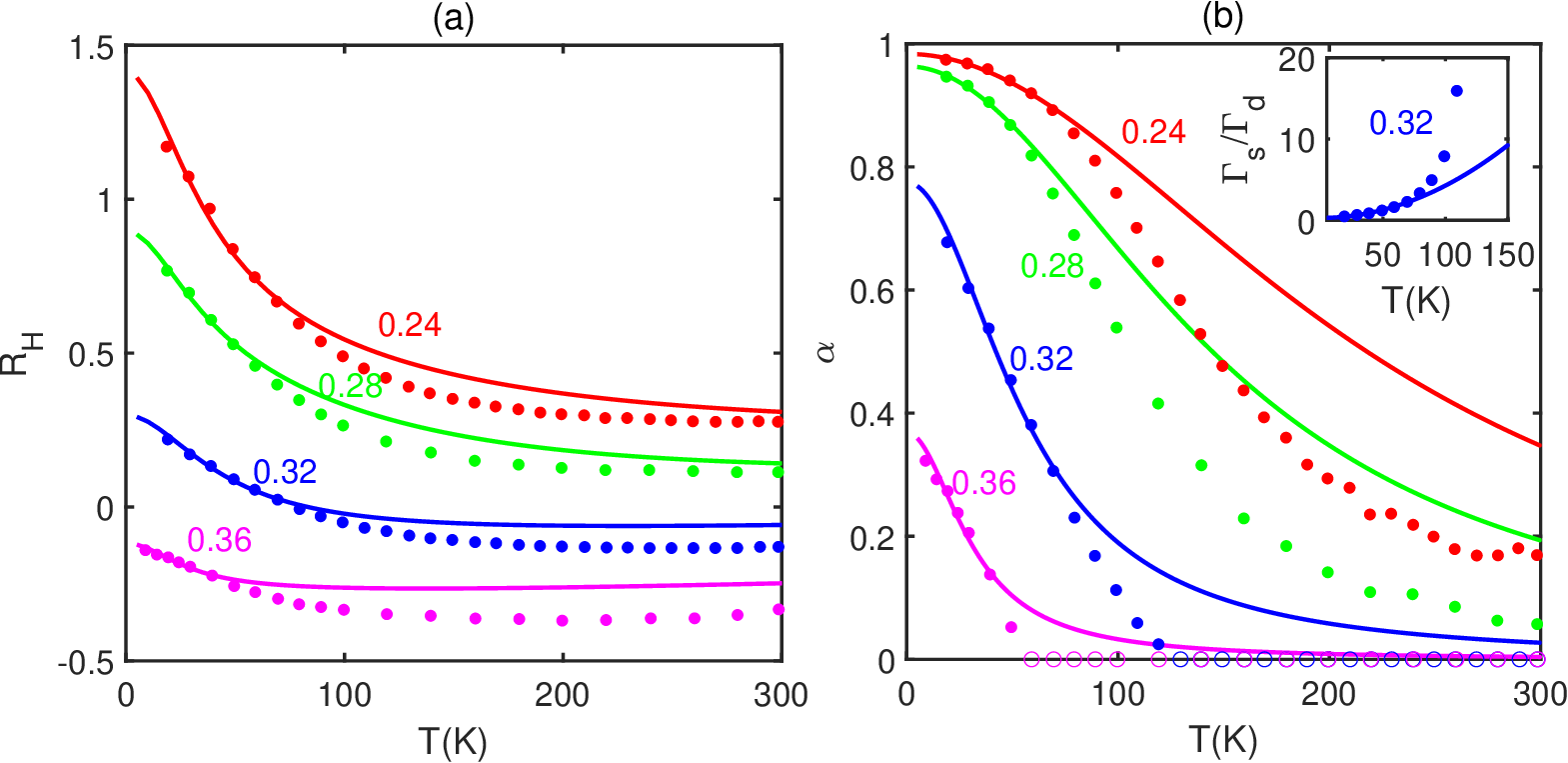}
\caption{The experimental data \cite{Tsukada2006} of $R_H$ (in unit of $a^2c/e$)  versus temperature $T$, shown as dots in (a), are fitted point-by-point to extract the anisotropy ratio $\alpha$ shown in (b). The solid lines in (a) and (b) are fits to the symbols by assuming $\Gamma_s=\Gamma_{s0}+g T^2$. The doping level is indicated on the colored line. The inset in (b) plots ${\Gamma_s}/{\Gamma_d}$ versus $T$ at $x=0.32$.
The open circles in (b) indicate the temperatures at which the fit to the data (not shown) fails to obtain an $\alpha\in [0,1]$. }
\label{fig:rh-T}
\end{figure}


We now consider the temperature dependence of $R_H$.
The experimental data of $R_H$ \cite{Tsukada2006} in unit of $a^2c/e$ are shown as dots in Fig.~\ref{fig:rh-T}(a),
where $a=3.8{\rm\AA}$ and $c=6.6{\rm \AA}$ are lattice constants of LSCO \cite{Horio2018} and $e$ is the electron charge.
For each temperature and doping, we tune the anisotropy ratio $\alpha$ to fit the experimental data of $R_H$.
The fitted results of $\alpha$ are shown as dots in Fig.~\ref{fig:rh-T}(b).
At high temperature, we cannot fit the experimental data within a reasonable range $\alpha\in[0,1]$, which are represented by open circles.

To understand the temperature dependence in $R_H$ as well as in $\alpha$, we assume the isotropic part, $\Gamma_s$, receives Fermi-liquid correction from electron-electron interactions. It therefore decreases as the temperature is lowered. On the other hand, the anisotropic part, $\Gamma_d$, is caused by apical oxygen vacancies, hence is independent of the temperature. Therefore, the scattering anisotropy $\alpha=\Gamma_d/(\Gamma_d+\Gamma_s)$ is expected to increase as the temperature is lowered. As a result, the hole-like contribution to $R_H$ from nodal quasiparticles becomes larger relatively. This provides a mechanism for the upturn of $R_H$ as the temperature is lowered.

To see how this scenario works, we assume  $\Gamma_s=\Gamma_{s0}+gT^2$ and $\Gamma_d=\Gamma_{d0}$, correspondingly  $\alpha^{-1}=1+\Gamma_s/\Gamma_d$, to fit the symbols in both panels of Fig.~\ref{fig:rh-T}.
Note that the total scattering rate $\Gamma_s+\Gamma_d$ has no effect on $R_H$, but only their ratio $\Gamma_s\Gamma_d$ determines $R_H$ at each doping.
The fitted results of $R_H$ are presented as solid lines. We observe fair agreement between the dots and the solid lines at low temperatures. In particular, the inset of Fig.~\ref{fig:rh-T}(b) plots $\Gamma_s/\Gamma_d$ versus $T$ in the case of $x=0.32$. The $T^2$-dependence at low temperatures is obvious. We notice that the $T$-dependence of $\alpha$ has been pointed out in Refs.\cite{narduzzo2008violation,grissonnanche2021linear}. On the other hand, the upturn of $R_H$ becomes even more significant upon underdoping \cite{Hwang1994,Tsukada2006}, where the mechanism should go beyond the Fermi liquid picture because of the emergence of the pseudogap.

\begin{figure}
\includegraphics[width=\linewidth]{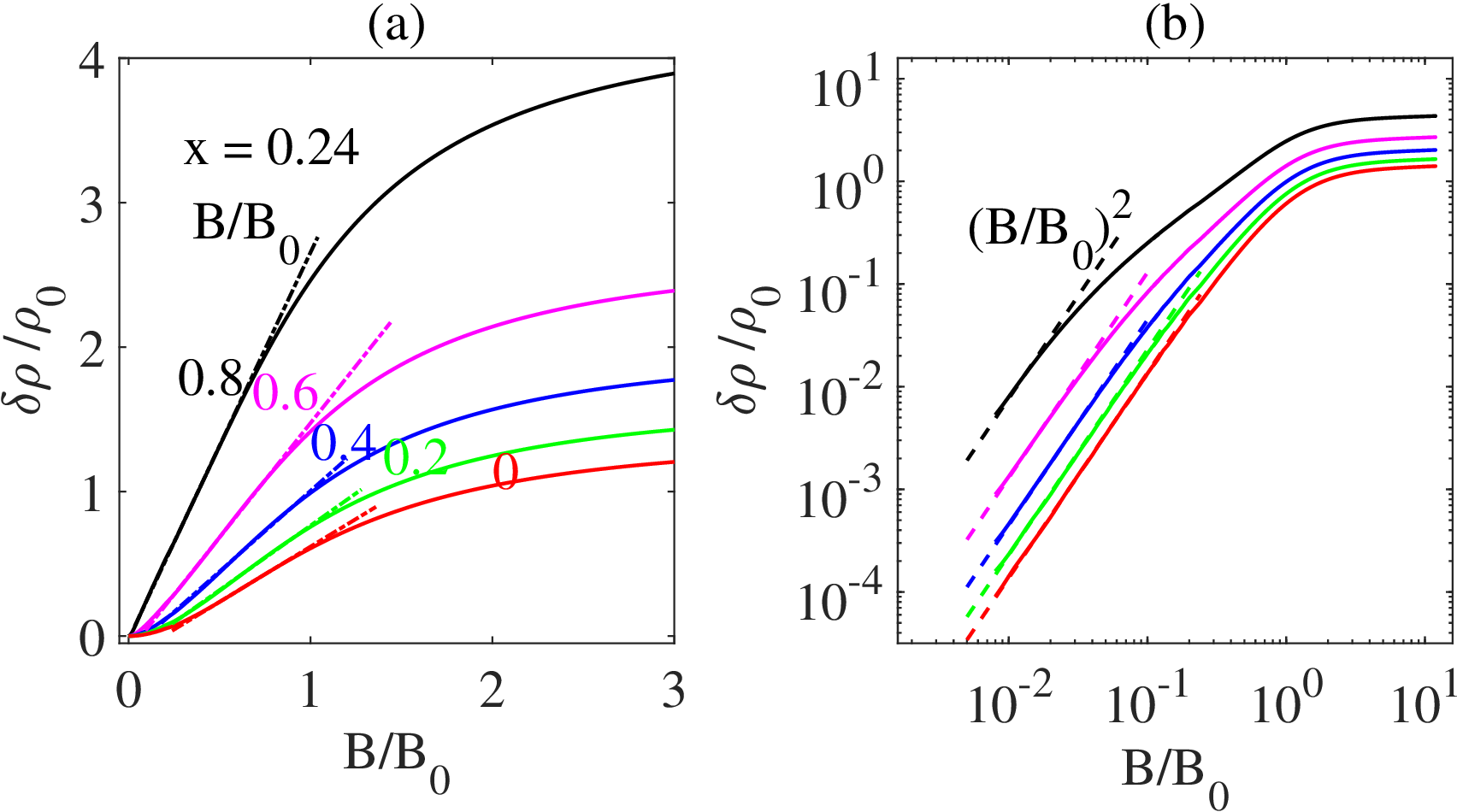}
\caption{(a) The magnetoresistivity ratio $\delta\rho/\rho_0$ are plotted versus $B$ at $x=0.24$ for $\alpha=0, 0.2, \cdots, 0.8$. The dashed lines show $B$-linear fits for the intermediate fields. (b) is the same as (a) but presented in log-log scale to reveal the $B^2$-behavior at low fields (highlighted by dashed lines). }
\label{fig:kohler}
\end{figure}

\subsection{Magnetoresistivity}

We now discuss the field dependence of the resistivity,  or magnetoresistivity. Previously, we found that the proximity to the VHS can lead to quasi-linear dependence of the resistivity $\rho$ in a sizable window of intermediate magnetic field $B$\cite{Mao2021}, in qualitative agreement with the experiment\cite{Giraldo-Gallo2018}. At that stage the simplest isotropic scattering was assumed. Here we examine further the effect of the anisotropic scattering $\Gamma_d$. In Fig.~\ref{fig:kohler}(a), the dimensionless magnetoresistivity ratio  $\delta\rho/\rho_0=[\rho(B)-\rho(0)]/\rho(0)$ is plotted versus the field for several values of $\alpha$. Fig.~\ref{fig:kohler}(b) is a replot of Fig.~\ref{fig:kohler}(a) in the log-log scale. These results show that the magnetoresistivity is quadratic in $B$ at small fields, quasi-linear in intermediate fields, and saturates at high fields. In comparison to the case of $\alpha=0$ (isotropic scattering), we see that a larger anisotropy reduces the $B^2$ regime on one hand, and enhances the saturation value in the large field limit on the other hand. Therefore, the intermediate crossover (quasi $B$-linear) regime is enlarged by a larger $\alpha$. If we assume that $\alpha$ becomes effectively larger as the temperature is lowered, as we discussed in the previous subsection, the above result implies the $B$-linear regime extends as the temperature is lowered, consistent with the experimental results \cite{Giraldo-Gallo2018,Cooper2009}. This adds salt to the conventional mechanism based on the cyclotron motion \cite{Mao2021,grissonnanche2021linear,Taillefer2022}, which should be compared to the more exotic picture of ``quadrature Planckian dissipation'' \cite{Zaanen2019,Ayres2021,Berben2022}.

The above result can also be related to the so-called Kohler's law, which says that the magnetoresistivity ratio $\delta\rho/\rho_0$ is a universal function of $B/\rho_0$.
This is true only if $\alpha$ does not change with temperature such that the data at different temperatures fall on the same line.
But in LSCO, since $\alpha$ grows as the temperature is lowered (see the previous subsection), the Kohler's law is expected to be violated more quickly at lower temperatures. On the other hand, a larger value of $\alpha$ also leads to a larger slope in  $\delta\rho/\rho_0$ as a function of $B$ in the intermediate linear regime (which actually extends to very small fields), see Fig.\ref{fig:kohler}(a). These features are consistent with the experiments \cite{Kimura1996,Vanacken2005}. Therefore, variation of the scattering anisotropy in temperature provides a plausible explanation of the violation of Kohler's law.

\begin{figure}
\includegraphics[width=0.75\linewidth]{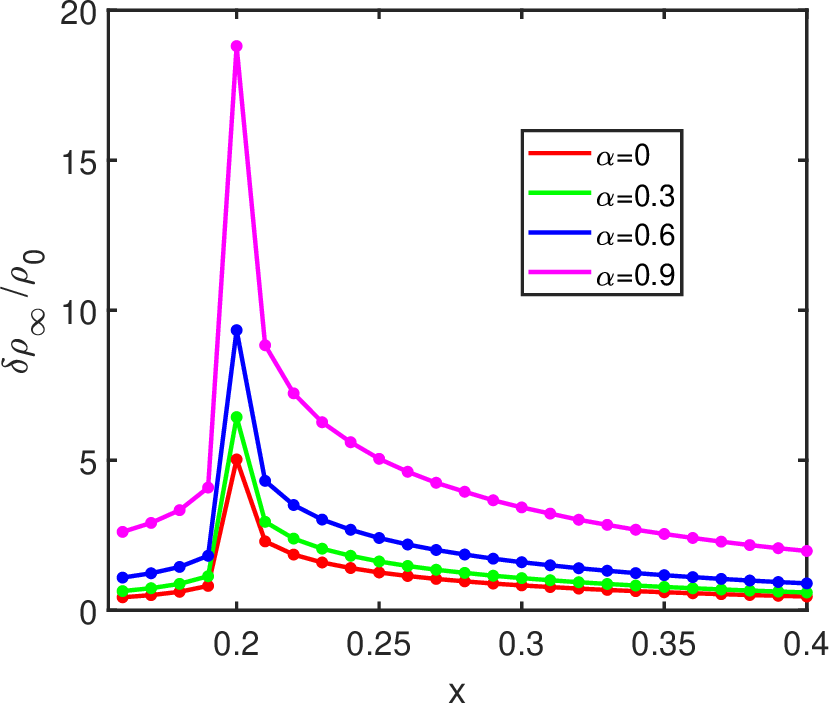}
\caption{Doping dependence of the saturated magnetoresistivity ratio $\delta\rho_{\infty}/\rho_0$ by fixing $\alpha=0$, $0.3$, $0.6$, and $0.9$, respectively.}
\label{fig:rhox}
\end{figure}

Finally, we examine the doping dependence of the magnetoresistivity. In Fig.~\ref{fig:rhox}, we plot the saturated magnetoresistivity ratio $\delta\rho_\infty/\rho_0$ versus $x$ for some specific values of $\alpha$. We see that it increases as the VHS doping is approached. This behavior is caused by the doping dependence of the cyclotron mass, which diverges at the VHS, as shown in our previous work \cite{Mao2021}. Here, we see that the magnetoresistivity ratio is further enhanced by the scattering anisotropy. The magnetoresistivity ratio peaks at the VHS, and the global scale clearly grows for larger $\alpha$ (at lower temperature in our picture). These predictions are to be checked in experiments. Our results may also shed light on the experiments \cite{Kimura1996, Vanacken2005, Cooper2009, Giraldo-Gallo2018}, where the magnetoresistivity ratio was found to increase at lower temperatures, although the peak at the VHS has not been observed (possibly because the VHS is smeared by the dispersion in $k_z$, or because the magnetic field is not strong enough).

\section{Summary}
In this work, we studied the effect of the anisotropic scattering $\Gamma_d$ on the Hall coefficient and magnetoresistivity in overdoped LSCO. Our study is based on well-defined quasiparticles in cyclotron motions under the magnetic field. We find that combining the anisotropic scattering, this picture can explain a handful of experimental observations simultaneously: the upturn of $R_H$ at low temperatures \cite{Hwang1994,Tsukada2006}, initial drop of $R_H$ in magnetic field\cite{Balakirev2009,Collignon2017}, linear magnetoresistivity in a window of intermediate field\cite{Giraldo-Gallo2018}, violation of Kohler's law \cite{Kimura1996,Vanacken2005}, and the temperature (and doping) dependence of the magnetoresistivity ratio \cite{Kimura1996,Vanacken2005,Cooper2009,Giraldo-Gallo2018}. These consistencies indicate the Fermi liquid picture is applicable in a substantial range of overdoped LSCO.

A key ingredient in our discussion is the anisotropic scattering, which we assume to follow from apical oxygen vacancies in overdoped LSCO \cite{Kim2017} {and more details can be found in Ref.~\onlinecite{Wang2022}}.
We should also point out that the functional form of the anisotropic scattering has been invoked in earlier literatures, but the origin is quite different. Ioffe and Millis suggested the so-called ``cold spot'' model with the scattering rate $\sim\cos^2(2\theta)$ caused by dynamic pair fluctuations to explain the optical conductivity in optimally and underdoped regions\cite{Ioffe1998}. Abrahams and Varma also proposed a scattering rate $\sim v_F^{-1}(\theta)$ \cite{Varma2001} from the forward scattering caused by interlayer impurities. Signature of anisotropic scattering appears ARPES \cite{Abrahams2000, N.Koshizuka2000, H.Raffy2005, Yoshida_2007, J.Mesot2008, chang2013anisotropic} and magnetotransport \cite{abdel2006anisotropic, grissonnanche2021linear,Taillefer2022} for both overdoped and underdoped samples.
{Recently, different out-of-plane dopant impurities, such as Sr-atoms or farther O-vacancies, are also proposed to support anisotropic scatterings via {\it ab initio} calculations \cite{dft2022}.}
Our results call for careful analysis of the effect of different anisotropic scatterings not yet covered in this work.

D. W. thanks Congjun Wu for early collaborations on this topic and Jie Wu for helpful discussions about experimental details. This work is supported by the National Natural Science Foundation of China (under Grant No. 11874205, No. 12274205 and No. 11574134).

\bibliography{vhs_chambersaniso}

\end{document}